\documentclass[ijodl]{svjour}

\usepackage{times}
\usepackage{graphicx}

\newcommand{\doi}[1]{{\footnotesize{\textsf{doi:#1}}}}
\newcommand{\url}[1]{{\footnotesize{\textsf{#1}}}}

\renewcommand{\t}[1]{\textit{#1}}
\renewcommand{\r}[1]{{\small\textsf{#1}}}
\newcommand{\rc}[1]{{\footnotesize\textsf{#1}}}
\newcommand{\onecolumnfigure}[1]{\begin{center}\resizebox{\hsize}{!}{#1}\end{center}}
\newcommand{\twocolumnfigure}[1]{\begin{center}#1\end{center}}

\usepackage[dvips]{color}
\definecolor{xmldecl}{rgb}{0.6,0.0,0.6}
\definecolor{xmlns}{rgb}{0.9,0.3,0.3}
\definecolor{xmlel}{rgb}{0.3,0.3,1.0}
\definecolor{xmlatt}{rgb}{0.5,0.1,0.1}
\definecolor{content}{rgb}{0.0,0.0,0.0}
\renewcommand{\i}{\hspace*{1em}}

\begin{document}

\title{Pathways: Augmenting interoperability across scholarly repositories}
\author{Simeon Warner$^1$, Jeroen Bekaert$^2$, Carl Lagoze$^1$,  
Xiaoming Liu$^3$, Sandy Payette$^1$, Herbert Van de Sompel$^3$.}
\institute{%
$^1$ Computing and Information Science, Cornell University, Ithaca, NY 14853, USA\\
%\{simeon,lagoze,payette\}\@cs.cornell.edu
$^2$ Ghent University, Faculty of Engineering, Jozef Plateaustraat 22, 9000 Gent, Belgium\\
%jeroen.bekaert\@ugent.be
$^3$ Research Library, Los Alamos National Laboratory, Los Alamos, NM 87545, USA%
%\{liu\_x,herbertv\}\@lanl.gov%
}
\date{ } %Received: date / Revised version: date}
\authorrunning{Warner \textit{et al.}}
\titlerunning{Pathways: Augmenting interoperability across scholarly repositories}
\maketitle

\begin{abstract}
In the emerging eScience environment, repositories of papers, datasets, 
software, etc., should be the foundation of a global and natively-digital 
scholarly communications system. The current infrastructure falls far 
short of this goal. Cross-repository interoperability must be augmented 
to support the many workflows and value-chains involved in scholarly 
communication. This will not be achieved through the promotion of single 
repository architecture or content representation, but instead requires 
an interoperability framework to connect the many heterogeneous systems 
that will exist.

We present a simple data model and service architecture that augments 
repository interoperability to enable scholarly value-chains to be 
implemented. We describe an experiment that demonstrates how the proposed 
infrastructure can be deployed to implement the workflow involved in
the creation of an overlay journal over several different repository systems 
(Fedora, aDORe, DSpace and arXiv).
\end{abstract}

\section{Introduction}
\label{sec-intro}

The manner in which scholarly research is conducted is changing rapidly. This is most 
evident in Science and Engineering~\cite{ATKINS+03}, but similar revolutionary trends 
are becoming apparent across disciplines~\cite{WATERS03}. Improvements in computing and 
network technologies, digital data capture techniques, and powerful data mining 
techniques enable research practices that are highly collaborative, network-based, and 
data-intensive. Moreover, the notion of a unit of scholarly communication is changing 
fundamentally. Whereas in the paper world, the concept of a journal publication dominated 
the definition of a unit of communication, in the emerging eScience environment, units of 
communication are increasingly complex digital objects. The digital objects can 
aggregate datastreams with both a variety of media types and a variety of intellectual 
content types, including papers, datasets, simulations, software, dynamic knowledge 
representations, machine readable chemical structures, etc.. Repositories that host 
such complex digital objects are appearing on the network at a rapid pace.  

In the light of these profound changes, we envision the emergence of a natively digital 
scholarly communication infrastructure that has this wide variety of repositories as 
its foundation. This infrastructure would leverage the value of the digital objects
in the underlying repositories by making them accessible for use and re-use in many 
contexts. In this infrastructure, repositories are not regarded as static nodes 
in a scholarly communication system 
that are merely tasked with archiving digital objects that were deposited there by scholars. 
Rather, repositories are perceived as the building blocks of a global scholarly communication 
federation in whi\-ch each individual digital object can be the starting point of value 
chains with global reach. 

Implementation of this infrastructure brings up a variety of intriguing 
prospects and associated questions across the whole sociological-economical-legal-technical 
spectrum. In the Pathways project, a joint project between Cornell University 
and the Los Alamos National Laboratory, we are exploring the technical problem domain. 
We focus 
on identifying and specifying the fundamental components required to facilitate the 
emergence of a natively digital, repository-based scholarly communication system. 
Our research tries to find the appropriate level of cross-repository interoperability 
that will provide a sufficiently functional technical basis for the 
realization of the vision, and will stand a realistic chance of being implemented in existing 
and future repository systems.  

This work is important because the current level of cross-repository interoperability is 
inadequate to support advanced forms of communication. Different communities 
have followed their own perspectives on repository design, implementation and management as 
well as on digital object representation and identification. Current interoperability is 
provided mainly by support of the OAI-PMH~\cite{OAI-PMHv2} and its mandatory Dublin Core 
metadata format~\cite{DublinCore}. Realizing the vision will require significantly 
augmented cross-repository interoperability. 

The remainder of this paper is organized as follows. Section~\ref{sec-motivating-context}
presents several motivating scenarios, and then section~\ref{sec-related-work}
describes related interoperability work. The following sections then introduce ideas 
for a cross-repository interoperability framework that have resulted from the 
Pathways project. The proposed high-level requirements for participating 
repositories can be summarized as follows:

\begin{itemize}
\item Support for a shared \t{data model} for \t{digital objects}
  (section~\ref{sec-data-model}).
\item Support for a \t{surrogate} format that serializes the \t{digital object} in
  accordance with the \t{data model} (section~\ref{sec-surrogates}). 
\item Support for three core repository interfaces: \t{obtain}, \t{harvest} and 
  \t{put}, to allow dissemination and ingest of \t{surrogates} (section~\ref{sec-services}).
\end{itemize}

The proposed framework also requires a shared service registry that lists the 
network location of the core interfaces for participating repositories. 
Section~\ref{sec-registries} describes the service registry and possible format 
and semantic registries that would further empower the environment. 
In section~\ref{sec-experiments} we describe experiments to implement an 
overlay journal scenario (an example we will use repeatedly throughout this 
paper) using this framework over existing repositories.
Section~\ref{sec-future-work} presents plans for future work, 
and section~\ref{sec-conclusions} draws some conclusions.
A less technical exposition of these ideas is given in~\cite{VANDESOMPEL+06}.

%%%%%%%%%%%%%%%%%%%%%%%%%%%%%%%%%%%%%%%%%%%%%%%%%%%%%%%%%%%%%%%%%%%%%%%%%%%%%%%%%%%
\section{Motivating context}
\label{sec-motivating-context}

In order to gain insights into the characteristics of the desired interoperability 
framework, it is helpful to investigate 
scenarios that drive this need for augmented interoperability. We see two classes of 
cross-repository value-chains: rich cross-repository services and cross-repository 
scholarly communication workflows.

In the first class of cross-repository value chains, repositories are regarded as sources of 
materials that can be used in services with a reach beyond the boundaries of a single 
repository. Materials should be exposed by repositories in a manner that allows for 
the seamless emergence of rich and meaningful services.  Discovery services are 
an obvious example of this class, and, although support of the OAI-PMH has resulted in a 
suite of cross-repository discovery capabilities, their functionality remains limited. 
For example, imagine creating a special-purpose search engine that collects only machine-readable 
chemicals structures, expressed using the XML Chemical Markup Language (CML), contained 
in digital objects hosted by repositories worldwide.  The current interoperability 
environment provides neither the ability to expose digital objects at a repository 
interface in a manner that unambiguously reveals the digital object's constituent datastreams, 
nor the language to express their intellectual content type (e.g. chemical structure).  As a 
result, the creation of the cross-repository chemical search engine would currently be truly 
complex, and would involve numerous repository-specific trial and error procedures. 

Consider the case where monitoring agencies make semantically tagged data 
on Arctic sea ice available in interoperable repositories.
An automated alerting service might then be able to discover and use both
raw and processed data (with raw data provenance accurately indicated) 
to provide early warning of events such as the abrupt shrinkage in Arctic 
sea ice in 2005. The output might be a report, a new digital object, 
containing both static `snapshot' results and importing dynamically 
computed elements. Accurate versioning of datasets would allow 
readers to be made aware of later amended inputs and perhaps even to 
recompute the results included in the report based on machine-actionable 
descriptions of the transform and visualization service.
A newspaper article on the findings might reference the source reports 
allowing readers to delve into and understand the sources and the 
basis of the claims as far as their understanding permits. 

In the second class of cross-repository value chains, repositories are regarded as the basic 
building blocks of a digital communication system, and scholarly communication itself is seen 
as a global cross-repository workflow~\cite{VANDESOMPEL+04}. Digital objects contained in 
repositories are the subjects of the workflows, and are used and re-used in many contexts.  

Citation is probably the most obvious example of this. 
In today's scholarly communication system, citation is implemented by inserting textual 
information describing a cited paper at the end of the citing paper, either by just typing it, 
by copy/pasting it from a Web page, or by importing metadata from a personal bibliographic 
citation tool. Thus, citations that are included in a digital manuscript are purely textual and are not 
natively machine readable or machine actionable. As a result, various \textit{post-factum} approaches 
have been devised to connect citing paper to cited paper by means of hyperlinks in the Web 
environment~\cite{VANDESOMPEL03}. These approaches include fuzzy metadata-based citation 
matching~\cite{HITCHCOCK+02}, the DOI-based CrossRef linking environment~\cite{CrossRef}, 
and the OpenURL framework for context-sensitive linking~\cite{VANDESOMPEL+01}. The variable
quality of citation metadata, among other factors, means that none of these approaches is 
foolproof. Furthermore, it is challenging to imagine how these approaches would extend 
beyond conventional scholarly papers, into the realm of complex digital objects that contain datasets, 
simulations, visualizations and so forth. It is therefore intriguing to think about citation 
as the re-use of the cited digital object in the context of the citing digital object.

To understand this expanded view of citation, imagine being able to drag a machine readable 
representation of a digital object hosted by some repository, and to drop it into the citing 
object that, once finalized, is submitted into another repository. Now imagine being able to 
do the same for the citing object \textit{ad infinitum}. Assuming that the machine readable 
representations that are being dragged and dropped contain the appropriate properties, the 
result would be a natively machine-traversable citation graph that would span across repositories 
worldwide. With appropriate user tools this would not only be vastly more functional than 
current forms of citation, but also simpler to use and to manage.

Collectively, these scenarios lead to a number of high-level observations: 
\begin{description}
\item[\textbf{Long-term perspective}] ---\,\,
Scholarly communication is a long-lasting endeavor, and, as a consequence, a long-term 
perspective should inspire the thinking about a future digital scholarly communication 
infrastructure. Clearly, this yields requirements related not just to the longevity of 
repositories and their collections, but also to the interoperability framework. 
The framework should be defined with sufficient abstraction to allow implementation 
using different technologies as time goes by, and should not be tied to a specific 
type of identifier, but rather support all current and future identification systems.
\item[\textbf{Content-transfer is often unnecessary}] --- 
Most of the value chains illustrated in the above scenarios do not require
the transfer of all digital object content. Instead just a subset appropriate to the particular
value chain. For example, the citation scenario requires only the transfer of the bibliographic
metadata of the cited paper, whereas the search engine scenario only requires the transfer of the
chemical formula. Full content-transfer as required for repository mirroring is just one of many use
cases that should be enabled by a desired solution.
\item[\textbf{Fine grained identification}] ---
Identifiers of journal articles, such as DOIs,
are typically repository independent in the sense that copies of a paper with a given identifier
stored by multiple repositories share the same public identifier.  This level of identification
granularity is sufficient for citation purposes.  However, it becomes inadequate when trying to
record the chain of evidence for cross-repository value chains because these have a specific digital
object from a specific repository as their subject. This means that a finer level of identification
granularity is required than provided by the existing bibliographic infrastructure. 
\end{description}

%%%%%%%%%%%%%%%%%%%%%%%%%%%%%%%%%%%%%%%%%%%%%%%%%%%%%%%%%%%%%%%%%%%%%%%%%%%%%%%%%%%%%%
\section{Related work}
\label{sec-related-work}

Pathways is focused on defining a common \t{data model} and \t{service interfaces}. 
These are designed to enable re-use and re-combination of digital objects and 
their components, to facilitate workflows over distributed repositories, and to 
enable computation and transformation of digital objects with dynamic service 
linkages. A key aspect of this work is that it explicitly handles the notion of 
provenance or lineage when content is re-used.

A significant amount of work exists in the design and specification of data models 
for digital objects, and in the creation of XML representation formats to promote 
the interoperable transmission and exchange of digital objects.
XML representation formats include the Metadata Encoding and Transmission 
Standard (METS)~\cite{METS}, the MPEG-21 Digital Item Declaration Language~\cite{BEKAERT+03}, 
the IMS Content Packaging XML Binding~\cite{IMSCP}, and the 
XML Formatted Data Unit (XF\-DU)~\cite{XFDU}.
Many of these formats have been used to enable the transfer of digital assets among systems.
A notable example is the use of MPEG-21 DIDL in the transfer of the American Physical Society 
collections to Los Alamos National Laboratory~\cite{BEKAERT+05}. 

There is no doubt that multiple data models for complex objects exist and will 
continue to be favored by different communities. The challenge is to develop a simple 
and flexible overlay data model that does not depend upon asset transfer, and can 
accommodate the essence of these different content models, yet can provide a simple 
low-barrier entry point for interoperability among repositories.

The Content Object Repository Discovery and Registration/Resolution Architecture 
(CORDRA)~\cite{KRAAN+05} is similar to Pathways in its goal to provide an open, standards-based 
model for repository interoperability. However, CORDRA is primarily focused on enabling 
interoperability between learning object repositories via federated registries of metadata 
catalogs. Unlike Pathways, CORDRA is specified upon a landscape of authored metadata.
The Pathways data model is built upon a generic, graph-based abstraction that does not 
prescribe specific metadata other than small set of key attributes for objects.
While CORDRA offers support for retrieval of content, Pathways addresses 
both retrieval and write for complex objects among heterogeneous repositories. 
Finally, while a distributed name resolution system (e.g., the handle 
system) is a necessary architectural pillar in CORDRA, the Pathways 
identifier scheme does not depend on a shared digital object identifier resolution 
service shared by distributed repositories.

There are a number of other projects in the higher education community devoted to 
the goal of repository interoperability within service-based architectures. 
Similarly motivated work is being done with the EduSource Community Layer (ECL)~\cite{ECL}, 
the DLF Asset Action Experiments~\cite{DLF-ASSET-ACTION}, 
the Open Knowledge Initiative Open Service Interface Definitions (OSIDs)~\cite{OKIPROJECT}.
These projects each specify a middleware service layer to enable applications to be built 
over heterogeneous data sources. Pathways is distinguished in that it is focused on 
defining a minimal set of read/write services necessary to enable access and re-use 
of complex objects in a distributed, heterogeneous repository environment.
The intent of Pathways is to specify relatively lightweight services that are easy to 
deploy over existing architectures. At the same time, Pathways is motivated 
to provide a model that can record and exploit provenance relationships as 
content is re-used across different services.

The challenge of service-based repository interoperability is being taken up by many other 
communities, often with different definitions of the basic concept of a ``repository''. 
In terms of service interfaces and APIs, repository interoperability is being 
addressed both from an access perspective and an authoring perspective (i.e., write, put).
Many efforts are positioned around a limited view of ``content'', typically individual 
content byte streams (an image, a web page, a PDF document), or hierarchies of 
content byte streams with simple descriptors.

For example, there are many new services that position the web as both a readable and writable 
space, albeit in a limited manner.  Atom~\cite{ATOM} provides an API for an application 
level protocol for publishing and editing web resources.  It also provides an XML data 
format that can be used in both the syndication and authoring of content.  
The Amazon S3~\cite{AMAZON-S3} web service provides an interface 
to support reading and writing, ultimately providing an internet data storage service that is 
scalable, reliable, and fast.  SRW Search/Retrieve and Update~\cite{SRW-RU} defines 
a web-service interface for retrieving and updating metadata records.
Web-based Distributed Authoring and Versioning 
(WebDAV)~\cite{WebDAV} enables the web servers to be exposed as writable, in addition to readable, by 
providing an interface for uploading content using a file and directory paradigm. Each of these 
services share with Pathways the notion of simple web-based interfaces for creating and accessing 
content over the web.  However, a key distinction of Pathways is its focus on complex digital 
objects as units of content as compared to single-content byte streams (e.g., a file).  Another 
distinction of the Pathways work is that it is primarily intended to be an interoperability model 
for managed repositories, as distinguished from more nebulous storage services on the open web.

Pathways employs a graph or tree-based data model to overlay heterogeneous data sources, 
which is also the basis of several other efforts. 
Recently, JSR~170~\cite{JSR170} has garnered much attention. 
This is a specification of a Java-based API for interacting with heterogeneous 
``content repositories'' and repository-like applications in a uniform manner. 
The basic metaphor for interaction is that of a 
hierarchy of nodes with properties, where node properties can be either simple data types or 
binary streams. JSR~170 is positioned similarly to how JDBC is for relational databases. 
It is most useful for developing Java-based applications with a standard interface for 
connecting with content storage components (i.e., ``content repositories''). 
Since it is not web services oriented, it is not clear the impact it could have 
in providing interoperability among distributed institutional repositories, and 
in non-Java environments.

The Pathways framework is intended to be consistent with existing and
emerging web architecture principles and should be easily implemented using
existing web protocols and standards.  In considering the W3C recommendation
for the Architecture of the World Wide Web~\cite{JACOBS+04},
the Pathways framework has been influenced by the need for URI-based
identifiers for resources, the notion that resources can have one or more
``representations'', and that these representations can be sent or received via
simple protocols. Pathways is influenced by work in the semantic web
community, particularly the Resource Description Framework (RDF)~\cite{KLYNE+04}
as a model for expressing resources in a graph-oriented manner as resource 
nodes with property and relationship arcs.

%%%%%%%%%%%%%%%%%%%%%%%%%%%%%%%%%%%%%%%%%%%%%%%%%%%%%%%%%%%%%%%%%%%%%%%
\section{Digital objects, the Pathways Core data model, and surrogates}
\label{sec-model-and-surrogates}

\begin{figure*}[ht]
\twocolumnfigure{\includegraphics[height=8cm]{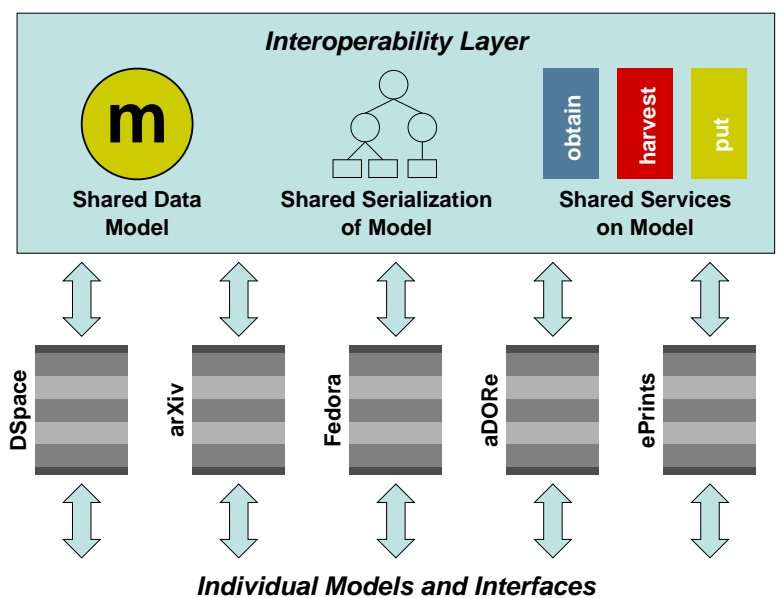}}
\caption{Interoperability layer over heterogeneous repositories.}
\label{fig-interop-layer}
\end{figure*}

The goal of our work on data models and interfaces is the creation of an
interoperability layer, as indicated in figure~\ref{fig-interop-layer}.
We expect that this layer will overlay 
data models and service interfaces that are distinct to individual repository
implementations. These repository-specific models and interfaces may provide
functionality outside and above the models and interfaces described here,
which are intended to represent the intersection (rather than union) of
individual repository features.

We use the following definitions:
\begin{description}
\item[\textbf{Digital object}] --- In the manner of the seminal Kahn and Wilensky
paper~\cite{KAHN+06} we use the notion of a digital object to describe compositions 
of digital information. This is purposely abstract, and is not tied to any
implementation or data model. The principal aspects of a digital object are
digital data and key-metadata. Digital data can be any combination and
quantity of individual datastreams, or physical streams of bits, and can
consist of nested digital objects. Key-metadata, at a minimum, includes an
identifier that is a key for service requests on the digital object at a
service point.
\item[\textbf{Data model}] --- We describe a data model, the Pathways Core, that
provides a formalization for overlaying digital objects on a network of
heterogeneous repositories and services.  We use UML to describe this data
model, but it could be described in other formalizations such as XML schema
or OWL.
\item[\textbf{Surrogate}] --- We use the term surrogate to indicate concrete
serializations of digital objects according to our data model.  The purpose
of this serialization is to allow exchange of information about digital
objects from one service to another and thus propagate them through value
chains.  We use RDF/XML for constructing our surrogates, because it is useful
for representing arbitrary sub-graphs.
\end{description}

The primary goal of the data model, and consequently the 
surrogates that represent it on the wire, is not asset or content-transfer. 
Rather we have designed the data model primarily as a framework that 
describes the abstract structure of information objects, and the properties 
of that abstract structure such as lineage, identity, and semantics.  The
linkage in the model from the abstract structure to the physical content is 
\textit{by-reference} rather than \textit{by-value} containment.

There are a number of good
reasons to not mandate asset transfer in the interoperability fabric. Full asset transfer
is necessary for only a subset of possible applications. One notable one is preservation
mirroring, and thus preservation frameworks such as the Reference Model for an Open
Archival Information System (OAIS)~\cite{OAIS02b} include the notion of information packages 
that imply full transfer of information units. 
Many other applications such as the overlay journal example described later
can be accommodated without the overhead of shipping all the bits between
repositories and services. By supporting by-reference content, the Pathways
Core model enables services to selectively decide when and if to dereference
and pull content into the service environment. This promotes the notion of
``service-tuned'' asset transfer, where each service can be configured to
respond to by-reference content in a manner appropriate to the context.

In a number of cases, full asset transfer is forbidden or undesirable.  For example, a
rights holder may be willing to allow inclusion of their asset in another context by means
of reference through a surrogate, but may be unwilling to transfer the datastream itself.  
Or, the rights holder might allow the asset transfer if the assets are placed in some
digital rights management (DRM) wrapper. 

Finally, static transfer of an asset may be undesirable in the case of dynamic information
objects, such as data sets derived from sensor networks.  We foresee a number of
applications in the scholarly domain where such dynamic objects are desirable, such as
astronomy publications that include the latest sky survey data.

\subsection{Pathways Core data model}
\label{sec-data-model}

The Pathways Core data model is based on the notion of a graph of abstract
\t{entities} with concrete \t{datastreams} as leaves. In this model, a 
digital object is a sub-graph rooted at an entity. The data model is 
designed to meet the following requirements:
\begin{enumerate}
\item It permits recursion for arbitrary levels of \t{entity} containment.
\item It provides an explicit link to the concrete representation, or component 
  \t{datastreams}, of the digital object.
\item It includes a notion of object identity that is independent of specific identifier 
  schemes.
\item It expresses lineage among objects, providing evidence of derivation and workflow 
  among objects.
\item It accommodates the linkage of semantic tags to information entities that extend the 
  functionality of format tags to the domain of complex, multi-part objects.
\item It allows the maintainer of the object to assert persistence of the availability 
  of a surrogate.
\end{enumerate}

\begin{figure*}[ht]
\twocolumnfigure{\includegraphics[height=9cm]{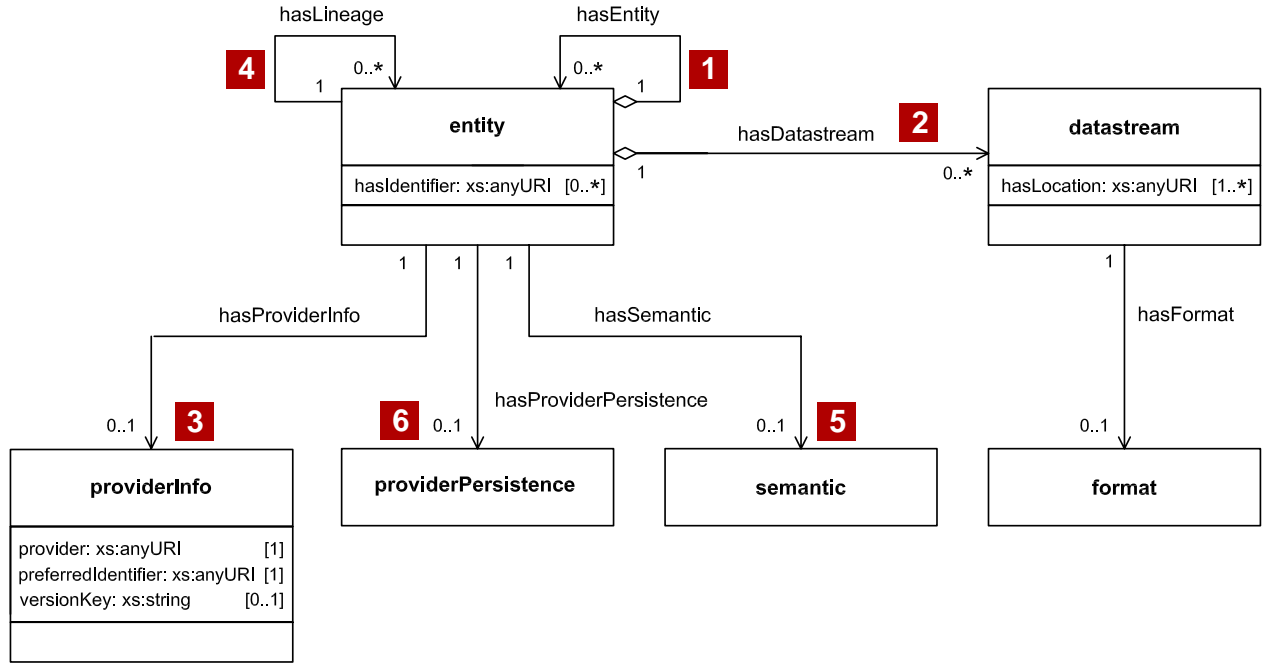}}
\caption{UML diagram of the Pathways Core data model with parts that fulfill particular 
requirements numbered.}
\label{fig-data-model}
\end{figure*}

A UML structure diagram of the Pathways Core is shown in figure~\ref{fig-data-model}.
The correspondence of features of the model to the requirements list above is 
indicated by the numbered properties. Each feature of the model is 
explained in more detail in the following sections. Our goal has been to find 
the minimal set of features necessary, the core properties. Certain uses or 
applications may require refinement of these relationships or the addition of 
new relationships, and we believe that such extensions can be added without 
breaking the core functionality.

\subsubsection{Entity recursion}

At the root of the Pathways Core is the notion of an \t{entity}. As shown in 
figure~\ref{fig-data-model}, this is the attachment point of a 
set of properties that associate the entity with its required and 
optional features. One property is \r{hasEntity}, which expresses recursive 
containment of entities. This maps to the Kahn/Wilensky~\cite{KAHN+06} notion
that digital objects can contain nested digital objects. An example of the 
utility of this recursive relationship is modeling of an overlay journal.
In this case, a top level entity could represent the journal itself, with
semantic, persistence, and identity attributes that correspond to the journal. 
A journal ``contains'' issues, which themselves may be entities, with 
associated properties. This recursion naturally continues, with issues 
``containing'' articles.

As indicated in figure~\ref{fig-entity}, an entity is an abstract concept, distinct 
from concrete datastreams described in the next section.  This abstract/concrete distinction is
fundamental to the model --- removing assertions of identity, persistence, lineage, and
semantics from individual physical manifestations of intellectual objects. This separation
of abstract and concrete properties (or attributes) is similar to that in the 
FRBR model~\cite{FRBR}.

\subsubsection{Concrete representation}

As indicated in figure~\ref{fig-data-model}, an entity can have several 
\r{hasDatastream} properties. 
The motivation for this is well-established in compound document formats such 
as METS, MPEG-21 DIDL, and Fedora FOXML, which allow a single object to have 
multiple datastreams with different media types (e.g., the availability of 
a scholarly paper in PDF, Word and TeX).

A datastream has both a \r{format} (e.g. a format registered in 
GDFR~\cite{ABRAMS+03} or PRONOM~\cite{DARLINGTON03}) and a \r{location}, a URL to 
request a dissemination of the datastream. The datastream association is 
intentionally \textit{by-reference} rather than \textit{by-value}, to avoid 
mandating asset transfer for the reasons given earlier.

\begin{figure}
\onecolumnfigure{\includegraphics{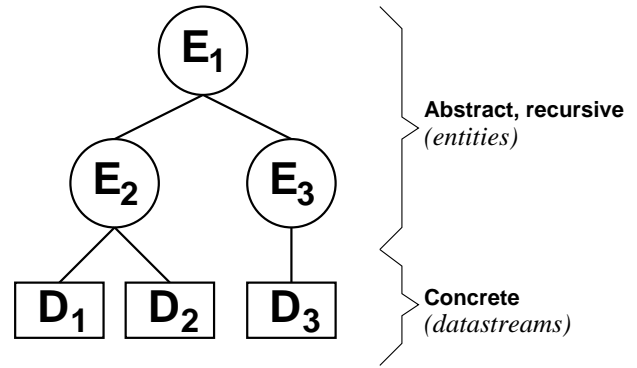}}
\caption{Entity recursion and concrete representation.}
\label{fig-entity}
\end{figure}
 
A typical digital object will contain one or more datastreams. 
The digital object represented in figure~\ref{fig-entity} comprises a 
top-level entity E$_1$, with sub-entities E$_2$ and E$_3$.
The entity E$_2$ has two datastreams, D$_1$ and D$_2$, which 
might be alternate expressions of the entity E$_2$. As semantic assertions
appear only at the level of the entity, both D$_1$ and D$_2$ 
are assumed to have any semantics expressed for E$_2$.
The entity E$_3$ has just a single datastreams D$_3$ and may 
thus be used to express semantics that apply just to D$_3$, as separate 
from the semantic of E$_1$. For example, E$_1$ might be a ``journal issue''
with two articles E$_2$ and E$_3$, and the article E$_2$ happens to be 
available in both PDF and Word formats.

\subsubsection{Identity}
\label{sec-identity}

We recognize the reality that one identifier technology will never dominate and
have thus incorporated two notions of identity. First, the \r{hasIdentifier} property
allows expression of URIs associated with a digital object, a DOI for example.
Second, the \r{hasProviderInfo} property introduces a relatively simple 
repository-centric identifier paradigm which permits precise identification
of digital objects in the particular repository, facilitating re-use and accurate
provenance records. This paradigm is not intended to replace existing identifier 
mechanisms or to interfere with future technologies in this area. 
Rather it is intended as a future-proof long-term scheme that can co-exist with 
other identifier mechanisms. 

The \r{hasProviderInfo} property has three components:

\begin{description}

\item[\r{provider}] --- The identity of the repository (i.e.; the service point 
providing access and ancillary services on the digital object).  
We assume that the participants in this infrastructure --- institutional 
repositories and the like --- have a commitment to of their repository
identity. Indirection via a repository identifier presumes some technology for
registering repositories and resolution to the location of their service interfaces.
Registries are discussed later, in section~\ref{sec-registries}.

\item[\r{preferredIdentifier}] --- The identity of the entity within the repository.  
This serves as the key for making service requests upon the digital object
at the service point defined by the repository (provider) identity.  As explained later in
this paper, the basic repository service is a request for a surrogate of the digital
object.  We expect, however, that a host of other services will evolve.  We emphasize that
the syntax, semantics, and resolution of the identity of the object is local to the
individual repository, rather than being global as in more ambitious identifier schemes.

\item[\r{versionKey}] --- This is a means of parameterizing a service request on an object
according to version semantics.  The intention here is to provide an opaque
hook into individual repository versioning implementations, rather than assuming or
imposing some universal cross-repository version schema.

\end{description}

Two copies of the same object in two different repositories may have the same identifier 
expressed via the \r{hasIdentifier} property. However, they will have different 
\r{providerInfo} because they are available from different repositories.

\subsubsection{Lineage}
\label{sec-lineage}

Isaac Newton wrote ``If I have seen further it is by standing on the shoulders of
Giants''~\cite{NEWTON1676}. In the face of massive changes in scholarship since Newton's
time, one constant is the evolution of scholarship, whereby new results are built on the
innovations of earlier scholars. We believe therefore that the interoperability
infrastructure must support the notion of lineage, natively linking entities to other
entities from which they are derived.

As shown in figure~\ref{fig-data-model}, entities in the model can link to other 
entities through the \r{hasLineage} relationship.  This linkage leverages the 
\r{hasProviderInfo} identity of the entity (or entities) from which the new
entity derives, thus allowing an entity to express its derivation from another 
entity and specifically state both the repository origin of the source object and 
its version semantics. Furthermore, since the model is recursive, entities can 
contain entities and the derivation of contained parts of objects can be similarly expressed.

\begin{figure}
\onecolumnfigure{\includegraphics[height=7cm]{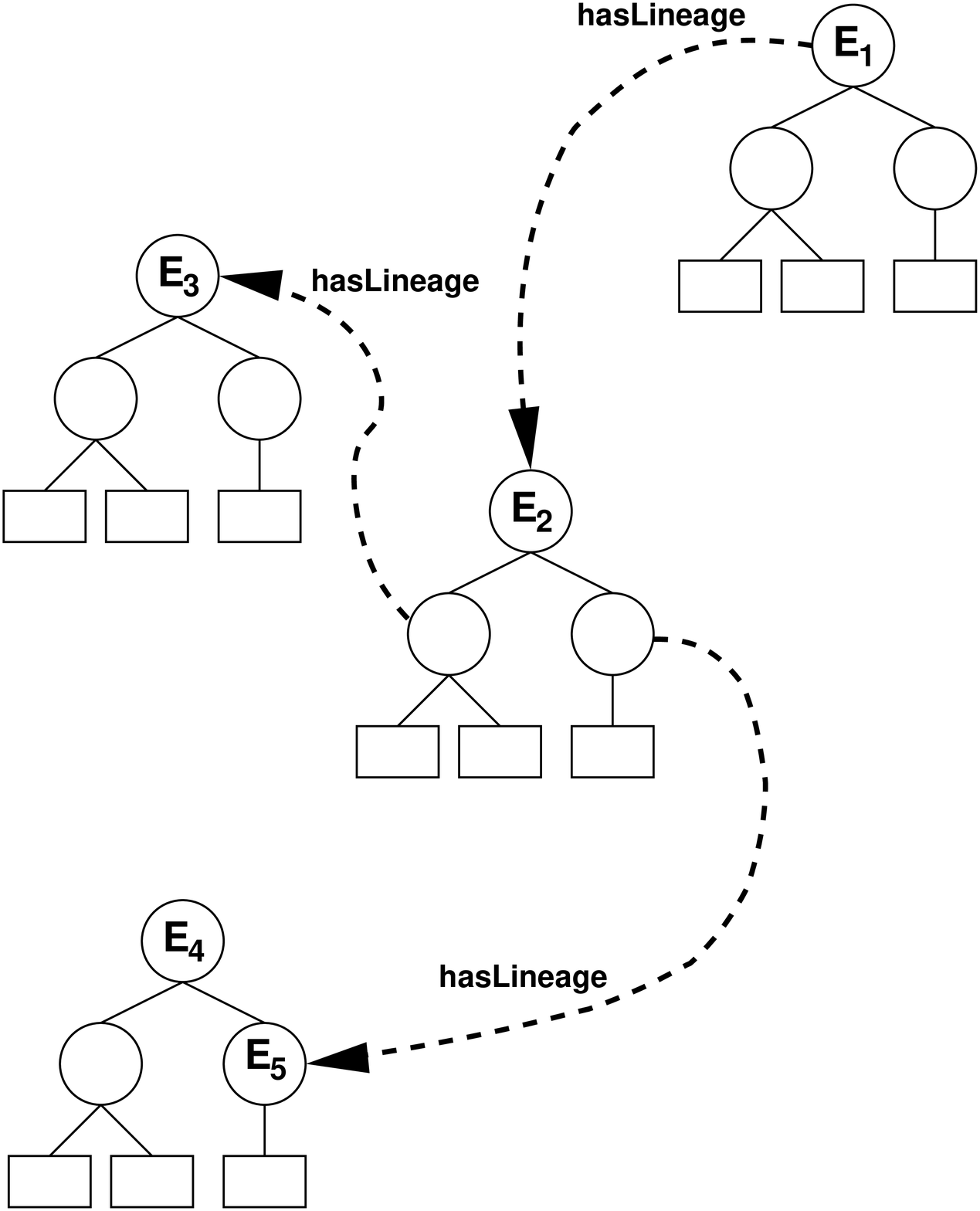}}
\caption{Relating entities by lineage.}
\label{fig-lineage}
\end{figure}

This lineage capability is illustrated in figure~\ref{fig-lineage}.
The entity labeled E$_1$ is derived from that labeled E$_2$. For example, 
E$_1$ may be translation of E$_2$ into a new language. E$_2$, as illustrated, 
contains sub-entities with respective derivations from E$_3$ and E$_5$. For example,
E$_2$ may be an issue of an overlay journal with articles that are edited versions 
of the preprints E$_3$ and E$_5$, where E$_5$ is itself a sub-entity of the 
preprint series E$_4$. These cases illustrate re-use at different granularities.

The result of these lineage links among entities at the interoperability layer is a 
web of \textit{evidential citation}. This graph indicates both the workflow origins of an 
information object --- the partial ordering of information objects from which it derives --- 
and also the curatorial heritage of the object --- the repositories and services responsible 
for its legacy. This new, uniquely networked and digital form of citation provides 
a finer level of identification than convention bibliographic citation. 
In the case that a repository has an object derived from an object in another
repository, there is a local choice as to whether the same object identifier
is used or a new one generated. This choice would be presumably be influenced 
by repository policy, community agreements and by the kind of value chain implemented. 
In either case, two observations can be made.
First, the \r{providerInfo} includes the \r{provider} which make the complete
identification unique and distinguished the objects. Second, the \r{hasLineage}
property of the derived entity provides and unambiguous link back to the original
entity. 

\begin{figure}
\onecolumnfigure{\includegraphics[height=7cm]{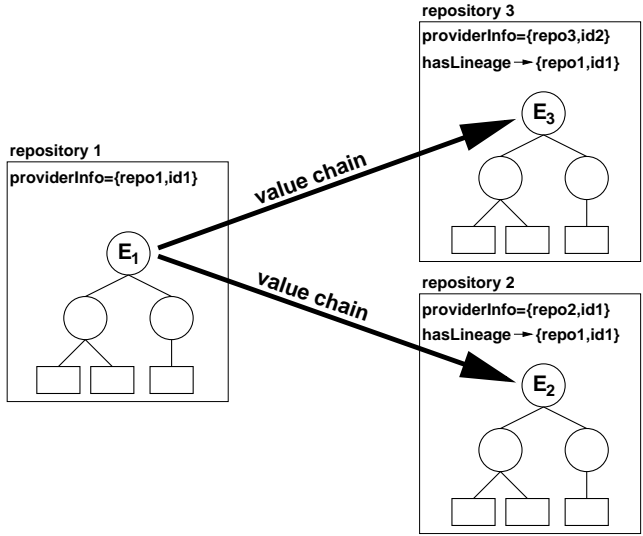}}
\caption{Identification and lineage of derived entities in different repositories.
The shorthand \{\rc{provider},\rc{preferredIdentifier}\} is used for \rc{providerInfo}, 
and \rc{versionKey} is omitted.}
\label{fig-lineage2}
\end{figure}

Both cases are illustrated in figure~\ref{fig-lineage2}, which shows 
the entity labeled E$_1$ taking part in value chains that result in new
entities, E$_2$ and E$_3$, in different repositories.  In all cases the 
entities are uniquely identified by the \r{providerInfo}, even though 
E$_2$ has the same \r{preferredIdentifier} as E$_1$. Also, both E$_2$ and E$_3$
indicate their lineage from E$_1$ with the \r{providerInfo} extracted from E$_1$.

We imagine that the \r{hasLineage} relationship is a superclass of the many types of
inter-entity derivation relationships that could be expressed. Thus, future evolution 
of the infrastructure might refine this relationship.

\subsubsection{Semantics}

We envision applications that need to know about the ``semantic'' composition of 
digital objects in addition to knowing the media-format types of the individual 
datastreams. A complex digital object might represent a ``dissertation'' or a 
``journal article'', each of which might have datastreams that are images, 
data sets, spreadsheets, or text in various formats. 

One particularly interesting application is service matching. The utility of automated
match of preservation services to information objects has been demonstrated by the PANIC 
work~\cite{HUNTER+04}. While PANIC demonstrates the utility of automation for individual 
datastreams based on media type, we would like to enable similar services over complex 
objects and based on intellectual content types.

The Pathways Core therefore associates the \r{hasSemantic} property with each entity. 
The target of this property is a URI specifying the semantic typing of the entity.  
Admittedly, no universal semantic registry exists at this time. However, the property
could be exploited by individual communities that develop local schemes, and 
later extended to more widespread use.

\subsubsection{Persistence}

The history of persistence of information artifacts, especially digital objects, is
riddled with examples of the gaps between intention, expectation, and reality. Despite our
best intentions to provide storage of and access to digital information ``forever'' (or
even a few months!), the realities of hardware failures, format rot, and mismanagement
frequently interfere. This must be considered in the design of any information
interoperability framework.

Therefore, we have taken a purposely modest approach to persistence that is oriented
towards surrogates and services over surrogates, rather than towards digital objects.  
The \r{hasProviderPersistence} property associated with an entity is a slot in which the 
repository can declare, by means of a URI, the longevity of its commitment towards providing 
services over the respective entity. The repository making this commitment is
identified as the \r{provider} in the entity's \r{hasProviderInfo}
property. Since the
core service in the interoperability fabric is the dissemination of a surrogate for the
entity, \r{hasProviderPersistence} indicates the level of commitment of the respective
repository to provide access to a surrogate for the entity. While there is clearly 
scope for subtle refinement of persistence declaration, at this point we propose a 
set of just two persistence declarations:

\begin{itemize}
\item The entity is transient and the repository makes no commitment
to providing services for it over time.
\item The entity is persistent and the
repository intends to respond to service requests for it over time. 
\end{itemize}

%%%%%%%%%%%%%%%%%%%%%%%%%%%%%%%%%%%%%%%%%%%%%%%%%%%%%%%%%%%%%%%%%%%%%%%%%%%%%%%%%%%%%%%%%%
\subsection{Surrogates and serialization}
\label{sec-surrogates}

An individual instance of the Pathways Core data model, a representation of an individual
digital object, is packaged and transmitted as a surrogate: a serialization that conforms
to the data model. We note possible terminological confusion here but have not found a word
with less baggage. By surrogate we mean a serialization that substitutes for the digital
object and must therefore reveal all essential characteristics, and is thus distinguished
from some arbitrary representation. The obtain and harvest interfaces (described
in sections~\ref{sec-obtain} and section~\ref{sec-harvest}) 
provide the means for clients to request a surrogate.  Similarly, a 
put request (described in section~\ref{sec-put}), which requests deposit of a digital 
object in a repository, contains a surrogate as a payload.

We have found that RDF~\cite{KLYNE+04} is a useful tool for modeling the graph-like
structure of information in the Pathways Core. We have done this by associating URIs with
the properties in the Pathways Core and similarly associating URIs with a number of
controlled vocabularies such as persistence, formats, and semantics that are the values of
Pathways Core properties. RDF modeling naturally led to the adoption of the 
RDF/XML syntax~\cite{BECKETT04} as the serialization syntax for Pathways Core surrogates. 
A fragment of an example of this syntax is shown in figure~\ref{fig-surrogate-xml}.

\begin{figure*}[ht]
\begin{minipage}[t]{\textwidth}\footnotesize\sf
\color{xmldecl}$<$?xml version="1.0" encoding="UTF-8"?$>$\color{content}\\
\color{xmlel}$<$rdf:RDF \color{xmlns}xmlns:core=\color{xmlatt}"info:pathways/core\#" \color{xmlns}xmlns:rdf=\color{xmlatt}"http://www.w3.org/1999/02/22-rdf-syntax-ns\#"\color{xmlel}$>$\color{content}\\
{\i}\color{xmlel}$<$core:entity \color{xmlns}rdf:about=\color{xmlatt}"info:pathways/entity/info\%3Asid\%2Flibrary.lanl.gov\%3Apathways/info\%3Adoi\%2F10.1016\%2Fj.dyepig.2004.12.010"\color{xmlel}$>$\color{content}\\
{\i\i}\color{xmlel}$<$core:hasSemantic \color{xmlns}rdf:resource=\color{xmlatt}"info:pathways/semantic/journal-article"\color{xmlel}/$>$\color{content}\\
{\i\i}\color{xmlel}$<$core:hasIdentifier$>$\color{content}info:doi/10.1016/j.dyepig.2004.12.010\color{xmlel}$<$/core:hasIdentifier$>$\color{content}\\
{\i\i}\color{xmlel}$<$core:hasProviderPersistence \color{xmlns}rdf:resource=\color{xmlatt}"info:pathways/persistence/persistent"\color{xmlel}/$>$\color{content}\\
{\i\i}\color{xmlel}$<$core:hasProviderInfo$>$\color{content}\\
{\i\i\i}\color{xmlel}$<$core:providerInfo$>$\color{content}\\
{\i\i\i\i}\color{xmlel}$<$core:preferredIdentifier$>$\color{content}info:doi/10.1016/j.dyepig.2004.12.010\color{xmlel}$<$/core:preferredIdentifier$>$\color{content}\\
{\i\i\i\i}\color{xmlel}$<$core:provider$>$\color{content}info:sid/library.lanl.gov:pathways\color{xmlel}$<$/core:provider$>$\color{content}\\
{\i\i\i}\color{xmlel}$<$/core:providerInfo$>$\color{content}\\
{\i\i}\color{xmlel}$<$/core:hasProviderInfo$>$\color{content}\\
{\i\i}\color{xmlel}$<$core:hasEntity$>$\color{content}\\
{\i\i\i}\color{xmlel}$<$core:entity\color{xmlns}rdf:about=\color{xmlatt}"info:pathways/entity/info...(shortened)...lanl-repo\%2Fssm\%2Fdoi-10.1016\%2Fj.dyepig.2004.12.010"\color{xmlel}$>$\color{content}\\
{\i\i\i\i}\color{xmlel}$<$core:hasSemantic \color{xmlns}rdf:resource=\color{xmlatt}"info:pathways/semantic/bibliographic-citation"\color{xmlel}/$>$\color{content}\\
{\i\i\i\i}\color{xmlel}$<$core:hasIdentifier$>$\color{content}info:lanl-repo/ssm/doi-10.1016/j.dyepig.2004.12.010\color{xmlel}$<$/core:hasIdentifier$>$\color{content}\\
{\i\i\i\i}\color{xmlel}$<$core:hasProviderPersistence \color{xmlns}rdf:resource=\color{xmlatt}"info:pathways/persistence/persistent"\color{xmlel}/$>$\color{content}\\
{\i\i\i\i}\color{xmlel}$<$core:hasProviderInfo$>$\color{content}\\
{\i\i\i\i\i}\color{xmlel}$<$core:providerInfo$>$\color{content}\\
{\i\i\i\i\i\i}\color{xmlel}$<$core:preferredIdentifier$>$\color{content}info:lanl-repo/ssm/doi-10.1016/j.dyepig.2004.12.010\color{xmlel}$<$/core:preferredIdentifier$>$\color{content}\\
{\i\i\i\i\i\i}\color{xmlel}$<$core:provider$>$\color{content}info:sid/library.lanl.gov:pathways\color{xmlel}$<$/core:provider$>$\color{content}\\
{\i\i\i\i\i}\color{xmlel}$<$/core:providerInfo$>$\color{content}\\
{\i\i\i\i}\color{xmlel}$<$/core:hasProviderInfo$>$\color{content}\\
{\i\i\i\i}\color{xmlel}$<$core:hasDatastream$>$\color{content}\\
{\i\i\i\i\i}\color{xmlel}$<$core:datastream$>$\color{content}\\
{\i\i\i\i\i\i}\color{xmlel}$<$core:hasFormat rdf:resource=\color{xmlatt}"info:pathways/fmt/pronom/1000"\color{xmlel}/$>$\color{content}\\
{\i\i\i\i\i\i}\color{xmlel}$<$core:hasLocation$>$\color{content}http://purl.lanl.gov/demo/adore-arcfile/00e682eb-a87eb27b0c79\color{xmlel}$<$/core:hasLocation$>$\color{content}\\
{\i\i\i\i\i}\color{xmlel}$<$/core:datastream$>$\color{content}\\
{\i\i\i\i}\color{xmlel}$<$/core:hasDatastream$>$\color{content}\\
{\i\i\i\i}\color{xmlel}...
\end{minipage}
\caption{Excerpt from a sample surrogate that serializes the Pathways Core in RDF/XML.}
\label{fig-surrogate-xml}
\end{figure*}

%%%%%%%%%%%%%%%%%%%%%%%%%%%%%%%%%%%%%%%%%%%%%%%%%%%%%%%%%%%%%%%%%%%%%%%%%%%%%%%%%%%%%%%%%%
\section{Repository interfaces: obtain, harvest and put}
\label{sec-services}

We have described the Pathways Core data model and a surrogate that serializes
the model. For these to enable repository interoperability, a set of essential 
services are required.  Three repository interfaces with the
following functions fulfill this need and are described below:

\begin{itemize}
\item An \t{obtain interface} which, in its most basic implementation, allows  
  the request of a surrogate for an identified digital object from a 
  repository.  
\item A \t{harvest interface} that exposes surrogates for incremental 
  collection or harvesting.
\item A \t{put interface} that supports submission of one or more surrogates 
  into the repository, thereby facilitating the addition of digital objects 
  to the collection of the repository.
\end{itemize}

\subsection{Obtain interface}
\label{sec-obtain}

Pathways defines an obtain interface that supports the request of services pertaining 
to an identified digital object within a repository. The simplest implementation 
of the obtain interface allows requesting a surrogate for an identified digital 
object from a repository. Such an interface
can be regarded as an identifier-to-surrogate resolution mechanism that 
resolves the \r{preferredIdentifier} of a digital object into a surrogate of 
that digital object.

The information needed to construct an obtain request is recorded in the \r{providerInfo}
property of the surrogate itself. The \r{providerInfo} is a triple consisting of the 
identifier of the repository that exposes the surrogate, the \r{preferredIdentifier} 
of the digital object and an optional \r{versionKey}. By using the identifier of the 
repository, the location of the obtain interface of the identified repository can 
be found by a look-up in a service registry (see section~\ref{sec-registries}). 
Once known, one can use the \r{preferredIdentifier} of the digital object (and the
optional \r{versionKey}) to obtain a surrogate using the repository's obtain
interface.

Higher levels of the obtain functionality have been explored theoretically by 
Bekaert~\cite{BEKAERT06}. Straightforward extension of the obtain concept allows 
the request of any supported service pertaining to an identified digital object. 
This includes the request of services pertaining to datastreams of the digital object. 
Possible examples are requests to obtain a surrogate of an identified article, 
requests to obtain a PDF datastream of that same article, requests to obtain an 
audio version of that article by applying a text-to-speech
service upon the PDF datastream, and so forth. Such services can be considered a superclass
of the basic obtain functionality described above, and do not have to be supported by all
repositories. Rather, such services would typically be supported by autonomous service
applications that overlay one or more repositories and use surrogates that are obtained
through interaction with the core obtain interface of the underlying repositories.

One technology that lends itself for implementing these obtain interfaces is the 
OpenURL Framework for Context-Sensitive Services~\cite{Z39_88_OpenURL}. 
The OpenURL standard originates from the scholarly information community where 
it was proposed a solution to the provision of context-sensitive reference 
links for scholarly works such as journal articles and books~\cite{VANDESOMPEL+01}. 
The initial standard was generalized to create the current NISO OpenURL Framework 
which describes a networked service environment, in which packages of
context information (\r{ContextObjects}) are used to request context-sensitive services
pertaining to a referenced resource. Each \r{ContextObject} contains various 
types of information that are needed to provide context-sensitive services. 
Such information may include the identifier of the referenced resource,
the \r{Referent}, the type of service that needs to be applied upon the 
\r{Referent} (the \r{ServiceType}), the network context in which the 
resource is referenced, and the context in which the service
request takes place.

In this way, the core obtain interface can be implemented as an 
OpenURL Framework Application. The \r{ContextObject} used in the obtain 
request conveys the following information:

\begin{description}
\item[A \r{Referent}] --- The digital object for which an obtain request is formulated. 
The \r{Referent} is described by means of its \r{preferredIdentifier}. 
\item[A \r{ServiceType}] --- The service that generates a surrogate of the identified 
digital object.
\end{description}

Beyond meeting our basic requirements, the OpenURL Framework has the 
following attractive properties: 

\begin{itemize}

\item it makes a clear distinction between the abstract definition of concepts and
their concrete representation and the protocol by which such representations are transported.
A \r{ContextObject} may be represented in many different formats and transported using many
different transport protocols, as technologies evolve. Yet, the concepts underlying the
OpenURL Framework persist over time. 

\item it does not make any presumptions about the identifier
namespace used for the identification of digital objects (or constituents thereof), and
hence, provides for an obtain interface that can be implemented across a broad variety of
repository systems.  

\item it allows information about the context in which the obtain request took
place to be conveyed. This information may allow delivery of context-sensitive service
requests. Of particular interest is information about the agent requesting the obtain service
(the \r{Requester}). This information could convey identity, and this would allow
responding differently to the same service request depending on whether the requesting agent
is a human or machine. Similarly, different humans could receive different disseminations
based on recorded preferences or access rights. The OpenURL Framework is purposely
generic and extensible, and would also support to convey the characteristics of a user's
terminal, the user's network context, and/or the user's location via the \r{Requester}
entity. Though, this type of context-related tuning may not be important when requesting
surrogates of digital objects, it may prove to be essential when requesting 
rich services pertaining to datastreams.

\end{itemize}

\subsection{Harvest interface}
\label{sec-harvest}

A harvest interface allows collecting or harvesting of surrogates of digital objects. 
In addition to the facility to harvest all the surrogates exposed by a repository, 
we believe it is necessary to provide a facility allowing some forms of selective harvesting. 
The simplest, and perhaps most useful, form of selective harvesting is to allow downstream 
applications to harvest surrogates only for those digital objects that were 
created or modified after a given date. This echoes the Open Archives Initiative Protocol 
for Metadata Harvesting (OAI-PMH)~\cite{OAI-PMHv2} with the same motivation: downstream
applications may need an up-to-date copy of all the surrogates from a repository in order to
provide some service, and incrementally harvesting surrogates of newly added or modified
digital objects is an efficient way to do this.

A harvest interface could be implemented using various technologies such as the OAI-PMH, 
RSS or Atom, or with a subset of more complex technologies such as SRU/SRW. The OAI-PMH is a well
established harvesting technology within the digital library community and allows aggregation
of metadata from compliant repositories using a datestamp-based harvesting strategy. Although
the OAI-PMH was first conceived for metadata harvesting, it can be used to transfer any
metadata or data format, including complex-object formats, expressed in XML according to an
XML Schema~\cite{VANDESOMPEL+04b}. The OAI-PMH is thus capable of providing the 
harvest functionality, and the ability to leverage existing OAI-PMH implementations
is a significant benefit.

To support the harvest interface, the underlying OAI-PMH interface must follow these
conventions:

\begin{itemize}

\item Each OAI-PMH item identifier must match the \r{preferredIdentifier} of the Pathways Core
digital object. This avoids the need for clients to record relationships between OAI-PMH
identifiers and digital object identifiers which can become complex in various aggregation
scenarios.

\item The OAI-PMH datestamps must be the datetime of creation or modification 
of the digital objects as discussed in~\cite{VANDESOMPEL+04b}.

\item It must provide a metadata format for surrogates as described in 
section~\ref{sec-surrogates}.

\end{itemize}

It is worth noting one possible issue. The OAI-PMH specification is bound to the HTTP protocol
and the XML syntax for transporting and serializing the harvested records. While this
approach proves to be satisfactory in the current technological environment, it may prove to
be inadequate as technologies evolve. If this work were to be tightly bound with the OAI-PMH
then an abstract model would need to be created. However, if OAI-PMH is used simply as one
possible technology to implement harvest functionality then it could later be replaced.

\subsection{Put interface}
\label{sec-put}

Pathways defines a put interface to promote interoperable transmission of
surrogates to one or more target digital repositories. As with the obtain interface, 
digital objects are expressed as surrogates. At the interface level, put 
operations are simple and unassuming. They can be understood as a
\textit{request for deposit} of a digital object. 
This distinguishes the put interface from similar operations found in other 
push-oriented services whose purpose is to facilitate upload of binary
content streams, or transfer of assets using community-specific content 
packages such as METS, IMS-CP, or MPEG-21 DIDL.

The put interface does not presuppose that target repositories conform to any particular
underlying storage scheme (e.g., hierarchal file system, a web server with directories,
relational database, etc.). Additionally, the put interface is neutral about the underlying
data model of target repositories. The only requirement is that digital objects be
represented as surrogates expressed in the Pathways Core, which is specifically designed to
transcend the particulars of heterogeneous data models. The graph-based nature of this model
provides the flexibility to support the submission of both simple and complex digital
objects.

The put interface, in combination with a surrogate, is intended as a means for transmitting
just enough information to enable a receiving repository to make decisions on how to process
a surrogate --- without anticipating or assuming an underlying repository's requirements
for ingest. Datastream content is expressed by-reference in the surrogate, via the
\r{location} property. With this constraint, a surrogate represents a ``shallow copy'' of 
a complex object since there is no transmission of raw content within the surrogate. As 
discussed earlier (section~\ref{sec-model-and-surrogates}), this constraint is motivated by the need for 
simplicity, and the desire to keep authentication and authorization concerns out of the functional 
definition of the put interface. Authentication, authorization and policy are expected to
be handled at service implementation layers.

Unlike the obtain and harvest interfaces, no protocol or technology stands out 
as an obvious implementation option for the put interface.

%%%%%%%%%%%%%%%%%%%%%%%%%%%%%%%%%%%%%%%%%%%%%%%%%%%%%%%%%%%%%%%%%%%%%%%%%%%%%%%%%%%%
\section{Registries}
\label{sec-registries}

The proposed interoperability framework requires at least one supporting infrastructure
component: a service registry to associate providers with services. Additional format and
semantic registries would significantly enrich the environment. No particular technical
implementation is implied by the use of the term registry, but rather the general ability to
record, share and retrieve terms of a controlled vocabulary alongside their associated
properties.

\subsection{Service registry}

\begin{figure*}
\twocolumnfigure{\includegraphics[height=8cm]{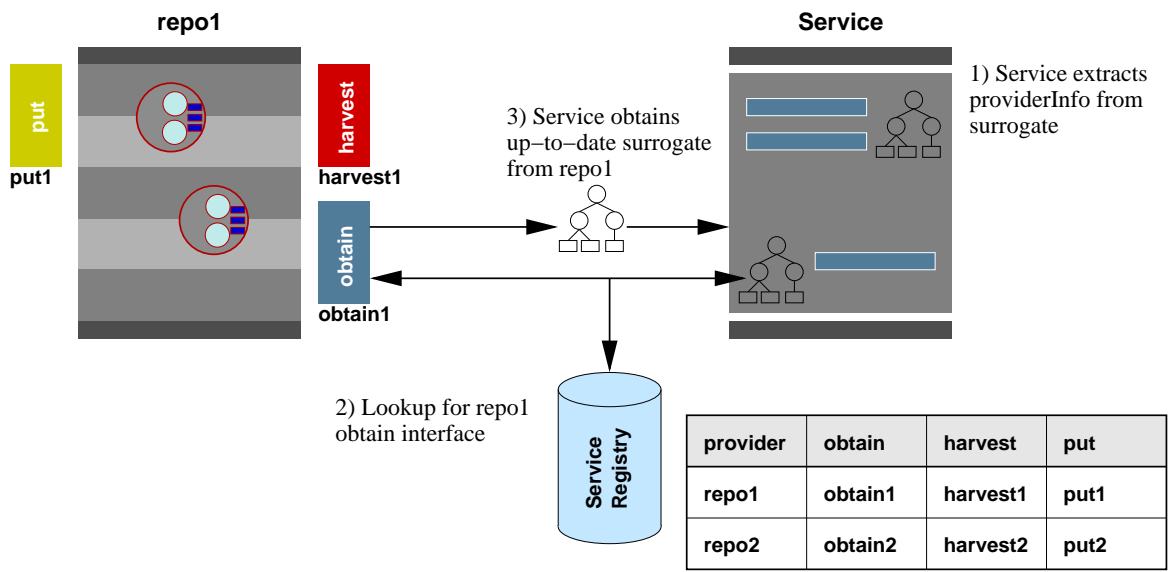}}
\caption{Use of the service registry.}
\label{fig-service-registry}
\end{figure*}

A service registry is fundamental to the framework, as it facilitates locating the core
service interfaces of participating repositories. This registry has the identifier of a
repository (\r{provider} from \r{providerInfo} in the Pathways Core) as its primary key, and it
minimally stores the actual network location of the obtain, harvest and put services, where
supported. Thus, given a surrogate with \r{providerInfo} (\r{provider}, \r{preferredIdentifier},
\r{versionKey}), it is possible for an application to use provider as a look-up key in the
service registry to retrieve the location of the core service interfaces for the repository
identified by provider. Once this information is available, actual service requests can be
issued against those interfaces. For example, in order to retrieve an up-to-date surrogate,
the application can issue an obtain request using the \r{preferredIdentifier} and optional
\r{versionKey} of the \r{providerInfo} as shown in figure~\ref{fig-service-registry}.
The use of a
registry permits repository interfaces to change their network location, allows different
services (including those not yet imagined) to be associated with a repository, and makes the
combination of repositories trivial.

It should be noted that, in contrast to other repository federation approaches such as 
CORDRA~\cite{KRAAN+05}, ADL-R~\cite{JEREZ+06}, aDORe~\cite{VANDESOMPEL+05}, 
and the Chinese Digital Museum Project~\cite{ChineseDSpace}, the
proposed framework does not require a registry of all digital objects in all contributing
repositories thereby allowing location of a digital object given its identifier. In the
proposed framework, a surrogate carries its self-identifying providerInfo, which, through the
intermediation of the service registry, allows location of the service interfaces of the
originating repository. This approach alleviates two major drawbacks inherent in the use of
digital object registries. First, given an identifier, how does one know that it is an
identifier of a digital object from repositories contributing to the federation, and hence
that a look-up in the federation's object registry is meaningful?  It seems that this
question can only be answered if all repositories in the federation share a common,
recognizable identifier scheme. This is a significant requirement, especially in light of the
considerations regarding the long-term horizon of desired solutions. Second, the scale of
object registries is several orders of magnitude larger than that of the proposed service
registry because the latter only contains an entry per repository, not per digital object.
The repercussions for operating the registry infrastructure are obvious.  

\subsection{Format and semantic registries}

While the service registry is essential for the operation of the proposed framework, two
other registries, while less fundamental, would significantly enrich the functionality of the
anticipated environment.

First, it is now widely recognized that repositories, especially in preservation
environments, must support more finely grained identification of digital media formats than is
provided by MIME types. A format registry that has the identifier of a digital format
as its primary key and that records various properties of the format have been proposed by
both the PRONOM~\cite{DARLINGTON03} and GDFR~\cite{ABRAMS+03} efforts. 
Format identifiers would be used
for the format property available at the datastream level of surrogates. Such a fine level of
format identification would, for example, enable rich format-based service matching as
explored in the PANIC~\cite{HUNTER+04} and aDORe~\cite{BEKAERT+04} efforts.

Second, automated object use and re-use would be enhanced by identification of the
intellectual content type of materials. A semantic registry that has the identifier of a
scholarly content type as its primary key and that records various properties of the content
type would support this. To facilitate syndicating, aggregating, post-processing and
multi-purposing magazine, news, catalog, book, and mainstream journal content, the PRISM
effort~\cite{PRISM_1_2h} has created such a vocabulary, but for materials typically used 
in a scholarly
context it is lacking, making the semantics registry probably more critical to pursue than
the format registry for which the MIME types can serve as a pragmatic stand-in. Semantic
identifiers would be used for the semantic property available at the entity level of
surrogates. Returning to the chemical search engine scenario, appropriate semantic
identification of an entity would allow an agent to recognize it as a machine readable
chemical formula, and thus choose to ingest the associated datastreams, the format of which
can also be precisely described.

%%%%%%%%%%%%%%%%%%%%%%%%%%%%%%%%%%%%%%%%%%%%%%%%%%%%%%%%%%%%%%%%%%%%%%%%%%%%%%%%%
\section{Experiments}
\label{sec-experiments} 

To test the ideas presented above, we created obtain, harvest and put 
services to disseminate and ingest surrogates from and to several different repository 
architectures: Fedora, aDORe, DSpace and arXiv. We then used these interfaces to support 
the assembly of a number of
articles from different repositories into a new issue of a hypothetical overlay journal.
Instead of relying just on the user interfaces of the participating repositories, we also
created a resource-centric search service using the harvesting infrastructure provided by the
harvest interfaces. To further enhance the demonstration, we combined these techniques with
Live Clipboard~\cite{LiveClipboard} technology to allow surrogates to be moved among 
repositories via the usual \textit{drag-and-drop} metaphor.
We first describe the two parts of this experiment, and then discuss implementation
issues and experiences.

\subsection{Harvesting journal articles to produce a resource-centric search service}
\label{sec-nutch}

\begin{figure*}
\twocolumnfigure{\includegraphics[height=9cm]{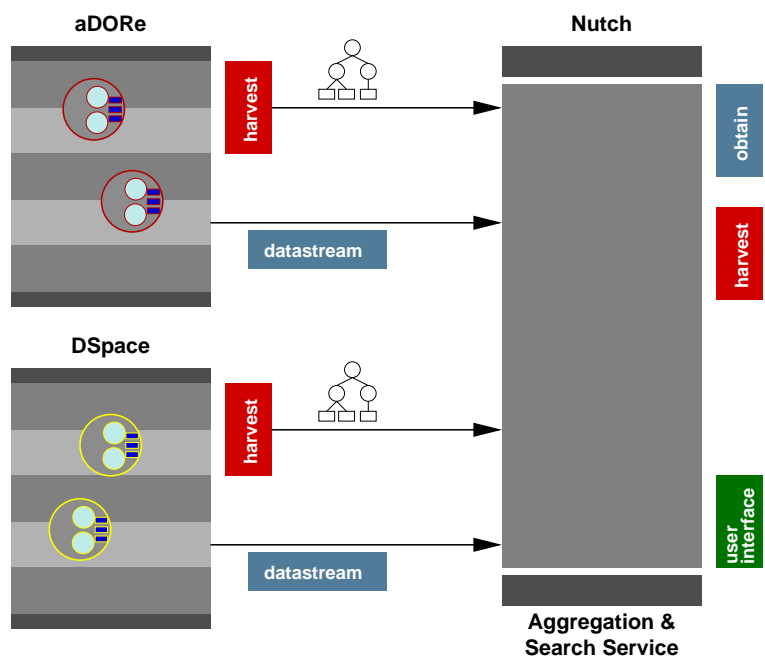}}
\caption{Use of Nutch to create a resource-centric search service over repositories
supporting the Pathways harvest interface.}
%%FIXME[hvds] - would be nice to remove harvest interface and add Live CLipboard links
\label{fig-nutch}
\end{figure*}

A number of projects have attempted to use OAI-PMH harvested metadata for the creation of
resource-centric or full-text discovery services. The principal problem is that
resources cannot be unambiguously located from the simple Dublin Core metadata exposed by
most OAI compliant repositories. This issue was discussed in detail in~\cite{VANDESOMPEL+04b}, 
where the use of complex object formats was proposed as a solution.
While skeletal compared with formats such as METS and MPEG-21 DIDL, the surrogates proposed
here also meet the requirements for resource harvesting. 
We implemented a search service
based on the Nutch~\cite{Nutch} crawler and search service. Instead of simply doing a web
crawl, surrogates were harvested from the participating repositories using the harvest
interface. Each surrogate was then introspected upon to select only those with the 
semantics ``journal-article'' (we agreed on a small ontology for these experiments). 
All the appropriate surrogates were examined to extract format and location information 
to dereference the datastreams. The datastreams were then fetched and indexed while retaining
their association with the surrogate. This process is illustrated in figure~\ref{fig-nutch}.

In addition to the usual links back to the source repository and content excerpt, the search
results display was augmented with a Live Clipboard icon allowing the surrogate to be 
copied into the copy/paste buffer on the user's computer, and thus easily passed to other
applications as described below.

\subsection{Creation of a new issue of an overlay journal}

\begin{figure*}
\twocolumnfigure{\includegraphics[height=6cm]{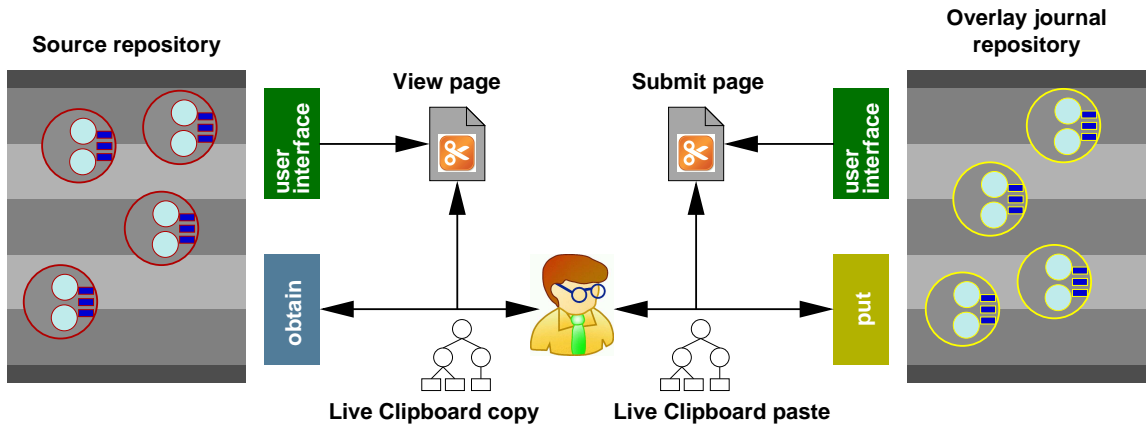}}
\caption{Addition of an article from a participating repository to an overlay journal using
Live Clipboard technology to copy a surrogate.}
\label{fig-overlay-journal}
\end{figure*}

The scenario we have referred to most frequently is the composition of a new issue of an
overlay journal from articles in different repositories. When combined with the search
service just described, this scenario demonstrates the use of all three repository interfaces
in a realistic scholarly value-chain. This scenario revolves around the editor of the overlay
journal, ``Ed''. The key data flows as Ed interacts with one source repository are shown
in figure~\ref{fig-overlay-journal}, and the complete sequence of actions required 
to create the new overlay journal issue are described below.

\begin{figure}
\onecolumnfigure{\includegraphics{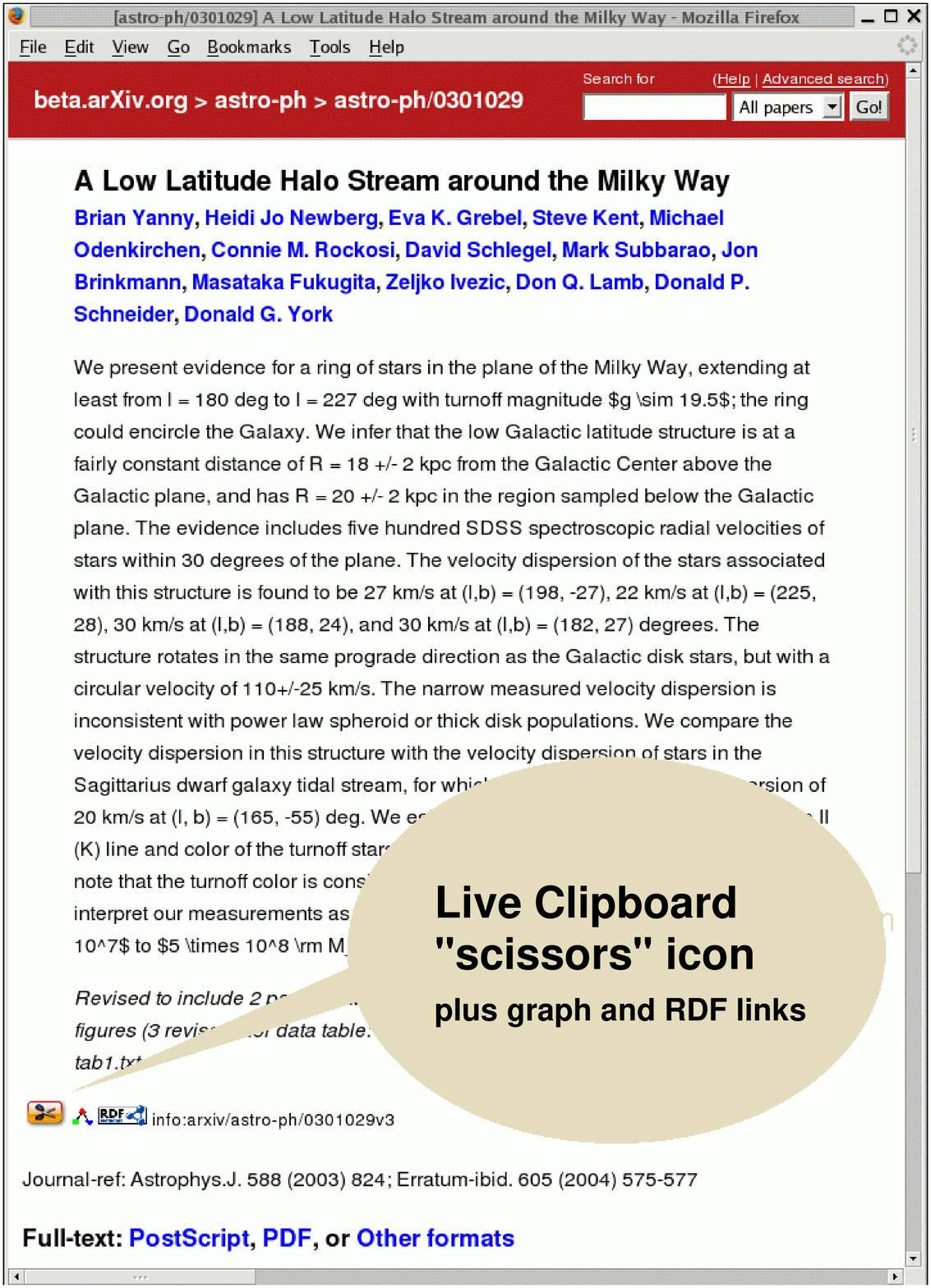}}
\caption{Screenshot of the arXiv wrapper page augmented with Live Clipboard links.}
\label{fig-browser-links}
\end{figure}

\begin{figure}
\onecolumnfigure{\includegraphics{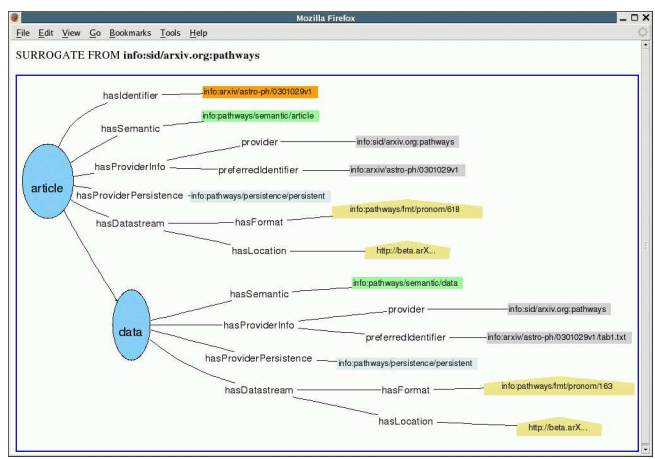}}
\caption{Screenshot of the graph visualization showing an article from arXiv that contains
a PDF file and a dataset.}
\label{fig-webdot-arxiv}
\end{figure}

\begin{enumerate}

\item \textbf{Select:} Ed applies whatever selection and review policies the overlay journal 
uses to decide which articles should be included in the new issue. Ed selects three articles; 
one each from arXiv, from an aDORe based repository, and from a DSpace repository.

\item \textbf{Obtain:} Consider first the selection of an article from arXiv. Ed navigates to the 
normal splash page for this article in whichever way is convenient, perhaps from Google, or from
arXiv's own interface. The splash page not only displays the usual metadata, links to
associate resources and links to the full-text, but also a Live Clipboard icon as shown in
figure~\ref{fig-browser-links}. By clicking
on this icon, the Live Clipboard JavaScript uses arXiv's obtain interface to get a surrogate
for the article which is stored in the copy/paste buffer of Ed's computer.

\item \textbf{Compose:} Ed then goes to the editorial web-interface for the overlay journal and pastes the
surrogate via the Live Clipboard JavaScript on that page. Behind the scenes, the surrogate is
put into the Fedora repository hosting the journal. Here it is a matter of local policy
whether the ingest mechanism simply stores the surrogate with references to included entities
and datastreams, or whether these are dereferenced and also ingested. For this demonstration
we chose to ingest only the structural information --- the entities --- which simulates a
``pure'' overlay which simply links to articles in trusted repositories (perhaps with
cyptographic signatures to guarantee that the original has not been altered). It would also
be possible for the ingest system to implement a deep-copy and duplicate all the datastreams of 
a digital object. Note that in a real system Ed would have to authenticate with the repository
for the overlay journal in order to be granted the privileges to put new content into the
journal, presumably any attempt to put content by a non-authenticated and or unprivileged
user would be denied.

\item \textbf{Complete composition:} A similar process is repeated for articles from DSpace and from aDORe. Here Ed uses the
search service described in~\ref{sec-nutch}. The search results show a Live Clipboard icon with
each result. By clicking on this icon, the Live Clipboard JavaScript uses the search service's
obtain interface to get a cached surrogate for the article which is stored in the copy/paste
buffer of Ed's computer. The overlay journal issue then has three articles queued.

\item \textbf{Submit:} When Ed is happy that all articles for the new issue are ready, the issue can be
created as an entity in its own right by clicking the ``Submit Issue'' button. When this is
complete, a surrogate for the new issue is available from the obtain and harvest interfaces
of the overlay journal repository and may be used by all the same services that interoperate
with the underlying repositories.

\item \textbf{Visualize:} To illustrate how other services can work within this 
framework, and to allow easy visualization of surrogates, we created an 
additional OpenURL-based service to visualize the surrogate graph using 
WebDot~\cite{WebDot}. 
Example output for an arXiv article containing an additional data datastream 
is shown in figure~\ref{fig-webdot-arxiv}. The OpenURL request simply includes the
providerInfo of the surrogate which is enough to enable the surrogate to be obtained, 
rendered as an image of a graph with links to sub-entities, datastreams and registry 
entries for format and semantic URIs. 

\end{enumerate}

Though only a demonstration, the process of compiling a new issue for an overlay journal
described above uses many interoperability features provided by the Pathways framework which
are simply not available in existing systems. By implementing this over several of the most
popular repository technologies we have demonstrated that this technology could readily be
deployed.

\subsection{Implementation of the obtain and harvest services}

The obtain interface is the simplest service and was the first that we implemented 
for each repository. By choosing to base it on OpenURL we were able to leverage 
existing OpenURL implementations for some of the repositories, simply adding 
another service identifier (\r{svc\_id}) for the obtain service. 
In implementing and obtain interface, one must work with the native data model
of the underlying repository. The key decisions arise in translation of digital
objects from their underlying representations to the Pathways Core model. Since the
Pathways Core model is flexible, this can be done in different ways, depending on
how much visibility vs. encapsulation of component parts is desired. Parts that
are made available for re-use should be modeled as entities with associated 
\r{providerInfo}.

All of the repositories used in this experiment already supported the OAI-PMH so the
implementation of the harvest interface was simply a matter of adding another ``metadata''
format for the Pathways Core surrogate to the existing OAI-PMH interfaces. 
It is a requirement of the OAI-PMH that all metadata formats be expressed in 
XML according to an XML Schema. Thus, we created surrogates using RDF/XML 
according to the W3C RDF/XML Schema~\cite{BRINKLEY+04}. Additionally, we 
agreed that, for convenience, all repositories would use a common 
\r{metadataPrefix=pwc.rdf} for this metadata format.

\subsection{Implementation of the put service}
\label{sec-put-implementation}

While the put interface is agnostic to the particulars of underlying repository technology
and models, any concrete implementation of a put service must be attentive to the specific
capabilities and limitations of the underlying repository architecture. There are many
questions that arise pertaining to how a put service interprets a surrogate and the
assumptions the service makes in interacting with a particular underlying repository.

To support the overlay journal experiment, a put service was developed to interact
with a Fedora repository. As no existing protocol already provided the required functionality,
a new REST-based service for Fedora was created for our experiments. 
The flexibility of Fedora made it well suited for accepting and ingesting both simple 
surrogates (single entity) and complex surrogates (a graph of entities). However, this 
flexibility provoked the realization that there are a number issues to be considered 
in implementing an effective put service, which we detail in the following sections.

\subsubsection{Identifiers and lineage}

The put interface, itself, imposes no requirement on a receiving repository in terms of how
it should deal with identifiers. However, whether the repository assigns new identifiers 
for surrogates ingested or not, it is important that there is a way to later determine
that an entity is a new instance of an existing entity. The means that the 
\r{providerInfo} of the original surrogate should be retained in the \r{hasLineage} 
property of the new entity as described in section~\ref{sec-lineage}.
Thus the \r{providerInfo} provides the basis for ``a chain of lineage'' across 
multiple distributed repositories where each repository represents the same entity 
or entities in different contexts.

\subsubsection{Ingesting hierarchies or networks of objects}

There are many cases when a put service will receive a surrogate that models a hierarchy or
graph of related entities. This presents a challenge in terms of determining an appropriate
ingest policy for how surrogates will be processed, and what kinds of digital objects will
ultimately be created in a receiving repository. 
 
When surrogates contain a hierarchy of entities, some assumptions must be made as to the
nature of the relationships of the entities in the hierarchy. Do parent-child relationships
of a hierarchy imply a part-whole composition? Is the presence or absence of each part
essential to the integrity of the whole? Alternatively, is the hierarchy to be interpreted as
a looser containment relationship, where the integrity of the whole is not compromised if its
parts are disassociated?

In our experiment, the put service assumed that any entity within a surrogate that 
contained \r{providerInfo} should be managed as its own digital object within the target 
Fedora repository. Thus, in the case of the journal overlay example, all
journal, issue, and article entities were to be represented as separate Fedora digital
objects with appropriate relationships asserted among them. Furthermore, any sub-entities of
article entities with \r{providerInfo} were also represented as digital objects in their own
right. In the experiments, there were article entities that were comprised of both a document
and datasets, and each of these was represented as a separate digital object.

The end result of creating a journal issue via the put interface, was the creation 
of a graph of related digital objects in the target Fedora repository. From a
management perspective, this modular and atomic arrangement can enable flexible
management of objects. For example, it is easier to discover and do something with all
dataset objects than it would be to find all types of objects that may encapsulate
datasets. From an access standpoint, each component is registered as a digital object with
its own public identifier and is available for re-use. This approach facilitated the 
ability to obtain the entire journal, or any sub-part down the hierarchy via the obtain 
service is a simple and generic manner.
It was not necessary to create special services to discover and extract entities that
were encapsulated within other objects. On the other hand, additional processes would be
required to implement facilities that depend on handling all the constituent parts of a given
digital object.

%%%%%%%%%%%%%%%%%%%%%%%%%%%%%%%%%%%%%%%%%%%%%%%%%%%%%%%%%%%%%%%%%%%%%%%%%%%%%%%%%%%%
\section{Future work}
\label{sec-future-work}

Much of the intellectual effort in this work has been to pare the Pathways Core data
model down to its essential components. Having successfully carried out some initial
experiments we intend to explore other scenarios, including those described in the
introduction, to see where additional richness is required. We envision extensions 
or refinements to relationships in the model. 
One example is the notion of entity containment to include more specific
semantics: distinction of part/whole vs. alternative or auxiliary; indication of
equivalence; and the 
notion of ordering or not. In the overlay journal example, how and where should the 
order of articles be expressed in a surrogate for the journal issue?

The hooks from URI expression of semantic information and format identifiers
open up tantalizing ties to current service matching work, the semantic web, and
to ontology-based reasoning. It seems likely that PRONOM and GDFR format
registries will coexist and be used by different segments of the scholarly
community. How can we use ontologies in systems that will ``understand'' the 
equivalences and differences between these specifications? How can notions of
generalization be applied to semantic information created in different contexts,
and how can we service match on compositional semantics? For example, if an
object contains a set of JPEG-2000 entities, should it be treated as a scanned
book or a photo album?

If we imagine a landscape of widespread re-use of digital objects, there will
undoubtedly be many copies and versions in different repositories and this provokes
a number of questions. 
When designing a put interface, how can one understand if a surrogate duplicates 
an existing digital object? When a duplicate is discovered as part of a put 
request, should the current object be replaced? Or should multiple versions 
of the digital object be managed? These are repository-specific decisions, 
but whatever is decided may have significant implications in collaborative 
scholarly workflows. How should surrogates be validated?

The experiments described here have been performed over repositories that
have primarily document content. The Pathways framework was conceived with
a rich environment of documents, data and other media-types in mind. 
Future work will involve collaboration with other repositories that
include significant data-repositories and other repository architectures.

%%%%%%%%%%%%%%%%%%%%%%%%%%%%%%%%%%%%%%%%%%%%%%%%%%%%%%%%%%%%%%%%%%%%%%%%%%%%%%%%%%
\section{Conclusions}
\label{sec-conclusions}

Our experiments successfully demonstrated the ability to move surrogates of
digital objects among repositories, and to re-use them in new contexts.
The proposed interoperability framework allowed us to show how the basic
workflow necessary to create a new issue of an overlay journal could be 
supported across heterogeneous repositories. The simplicity and generality 
of the Pathways Core data model allowed its use to create surrogates 
for digital objects held in Fedora, aDORe, DSpace 
and arXiv repositories, each with significantly different internal data 
models and architectures. Furthermore, by leveraging existing 
implementations of OpenURL resolvers and OAI-PMH interfaces for the 
repositories, it was remarkably easy to provide the dissemination 
(obtain) and harvest interfaces necessary for each repository to 
participate. The ingest (put) interface was implemented only for a Fedora 
repository and involved considerably more design decisions, many of which 
require further investigation to determine best practices.

These results are serving as the basis for further experiments with even 
more heterogeneous repository architectures, to include data repositories 
in particular. These experiments will implement other important value 
chains that are necessary to move toward the goal of the creation of a 
global scholarly communication system.

%%%%%%%%%%%%%%%%%%%%%%%%%%%%%%%%%%%%%%%%%%%%%%%%%%%%%%%%%%%%%%%%%%%%%%%%%%%%%%%%%%
\section*{Acknowledgements}

We thank Rob Tansley for implementing the Pathways Core data model and associated 
dissemination services in DSpace; Chris Wilper for work on Fedora, including the
put interface; and Lyudmila Balakireva for work on the WebDot service, and 
Zhiwu Xie for work on the the Live Clipboard infrastructure. 
This work was supported by NSF award number IIS-0430906 (Pathways).

\bibliographystyle{plainurl-sw2006-07-01.bst}
\bibliography{dl,pathways-ijdl-extra}

\end{document}